\documentclass[preprint,preprintnumbers,superscriptaddress,showkeys,showpacs,nofootinbib,aps]{revtex4}

\usepackage{amssymb}
\usepackage{epsfig}
\usepackage{amssymb}
\usepackage{epsf}
\usepackage{epsfig}
\usepackage{color}
\usepackage{ifpdf}
\usepackage{graphics}
\usepackage{longtable}

\newcommand{\nn}{\nonumber}
\newcommand{\ba}{\begin{eqnarray}}
\newcommand{\ea}{\end{eqnarray}}
\newcommand{\be}{\begin{equation}}
\newcommand{\ee}{\end{equation}}
\newcommand{\bd}{\begin{displaymath}}
\newcommand{\ed}{\end{displaymath}}

\newcommand{\plotangle}{270}

\def\desy{NIC, DESY, Platanenallee 6, D-15738 Zeuthen, Germany}

\def\HU{Institut f\"ur Physik, Humboldt-Universit\"at zu Berlin, D-12489 Berlin, Germany}
\def\JLab{Jefferson Lab, 12000 Jefferson Avenue, Newport News VA 23606, USA}
\def\Cyprus{The Cyprus Institute, P.O. Box 27456, 1645 Nicosia, Cyprus}
\def\kek{High Energy Accelerator Research Organization (KEK), Tsukuba 305-0801, Japan}
\def\Sokendai{School of High Energy Accelerator Science, The Graduate University for Advanced Studies (Sokendai), Tsukuba 305-0801, Japan}

\def\Cyprusuni{Departament of Physics, University of Cyprus, P.O. Box 20537, 1678 Nicosia, Cyprus}

\begin{document}

\preprint{KEK-CP-286}
\preprint{DESY 13-085}
\preprint{JLAB-THY-13-1736}
\preprint{SFB/CPP-13-30}
\preprint{HU-EP-13/24}

\title{Computing the hadronic vacuum polarization function 
by analytic continuation}

\author{Xu~Feng}
\affiliation{\kek}

\author{Shoji~Hashimoto}
\affiliation{\kek}
\affiliation{\Sokendai}

\author{Grit~Hotzel}
\affiliation{\HU} 

\author{Karl~Jansen}
\affiliation{\desy}
\affiliation{\Cyprusuni}

\author{Marcus~Petschlies}
\affiliation{\Cyprus}

\author{Dru~B.\ Renner}
\affiliation{\JLab}

\begin{abstract}
We propose a method to compute the hadronic vacuum polarization function
on the lattice at continuous values of photon momenta bridging between
the spacelike and timelike regions.
We provide two independent demonstrations to show that
this method leads to the desired hadronic vacuum polarization function 
in Minkowski spacetime. 
We show with the example of the leading-order QCD correction  
to the muon anomalous 
magnetic moment 
that this approach can provide a valuable alternative method 
for calculations of physical 
quantities where the hadronic vacuum polarization function enters. 
\end{abstract}

\keywords{hadronic vacuum polarization, muon g-2, lattice QCD}
\pacs{12.38.Gc, 14.60.Ef}

\maketitle

\section{Introduction}

In quantum field theory, the photon receives vacuum polarization corrections, which modify the electromagnetic coupling constant depending on the virtual photon momentum.
Quark-loop contributions to these corrections include the effects of the strong interaction through further gluon exchanges.
Since the strong interaction becomes nonperturbative
at low energies, the calculation of 
these hadronic vacuum polarization (HVP) effects requires 
a nonperturbative method to treat Quantum Chromodynamics (QCD).

The QCD corrections from the HVP appear in many 
physical observables, such as the running QED coupling,
the weak mixing angle, 
the $2P-2S$ Lamb shift of muonic hydrogen and the muon anomalous 
magnetic moment, $a_\mu$, \cite{Jegerlehner:2009ry,Hoecker:2010qn,Renner:2012fa}.
In fact, $a_\mu$ is a prime example where the HVP correction is very important 
since there is a 
discrepancy between the experimental measurement of this quantity 
and the Standard Model (SM)
prediction, and it is the HVP correction that constitutes the dominant theoretical
uncertainty. Since it is tempting to interpret 
the discrepancy as an indication for new physics beyond the 
SM, it is necessary to have a well-controlled QCD 
calculation of the HVP function before drawing any definitive conclusions.

At low energies, the running QCD coupling becomes large and perturbative QCD fails
to be a reliable tool for determining the HVP function. 
The established approach of using a dispersion relation together with experimental data for $e^+e^-$ scattering and $\tau$ decay depends on the experimental input and hence cannot be considered as an {\em ab initio} SM calculation.
It is therefore highly desirable to 
perform a nonperturbative calculation
of the HVP function from first principles using lattice QCD.
Indeed, for the quantities listed above, where the HVP function is an 
important ingredient, it has been shown that lattice QCD 
can provide a substantial, if not an essential, contribution; see Refs. \cite{Renner:2012fa,Bernecker:2011gh} for
a more detailed discussion.

In standard lattice calculations, the HVP function is calculated 
at discrete spacelike momenta 
by performing a Fourier transform of the Euclidean vector-vector correlation
function.
However, these lattice QCD computations suffer from a generic difficulty, namely, that low momenta are usually not directly accessible. For example, in the determination of the hadronic contribution
$a_\mu^{\rm hvp}$ to the muon anomalous 
magnetic moment, momenta close to 
$0.003$ GeV$^2$ dominate.
Accessing the HVP function at small or near-zero momentum is a general
concern relevant to
many observables to be computed in lattice QCD.
To address this problem,
efforts to approach the low
momentum region by using partially twisted boundary conditions have been undertaken~\cite{DellaMorte:2011aa}.
In addition, a number of approaches have been employed to describe 
the HVP function on the lattice. 
Here, well-motivated~\cite{Blum:2002ii,Gockeler:2003cw,Aubin:2006xv,Feng:2011zk,Boyle:2011hu,DellaMorte:2011aa} 
or even model-independent~\cite{Aubin:2012me} functional forms to fit the results  
computed at discrete lattice momenta have been used. They lead to 
smooth functions that describe well the desired momentum region,
including zero momentum, needed to 
obtain the renormalized HVP function. 
A step forward has been discussed in Ref.~\cite{deDivitiis:2012vs} 
where it was suggested to 
calculate the derivative of the relevant correlation function by using the sequential
source propagator method to obtain the zero 
momentum contribution. 

In this paper we 
discuss a more general approach
that allows us to compute the HVP function at continuous momenta,
both in the spacelike and timelike regions.
Following the ideas employed in Ref.~\cite{Ji:2001wha}
this will be achieved by starting with the Euclidean vector-vector correlation 
function and 
performing a Fourier transform only in the spatial directions
and an integration (in practice, a summation)
in the time direction with a factor of $e^{\omega t}$.
In this way, we are able to calculate the HVP function 
at very small spacelike momenta, covering also the zero 
momentum value and even extending in to the timelike region. 
We will provide two independent 
demonstrations that this leads to a sound and valid evaluation of the 
HVP function.  
In order to test whether this approach is practical and 
leads to small errors in the calculation, 
we apply this method to a realistic lattice calculation of the HVP function and the associated
$a_\mu^{\rm hvp}$.
We stress that we consider our approach as an alternative way to compute the HVP function in lattice QCD at small momenta.
It has the advantage 
of avoiding assumptions
on the analytic form used to describe the HVP function. 
However, as we will see below, it will not lead to an increased
precision in the calculation of physical quantities such as 
$a_\mu^{\rm hvp}$. Nevertheless, we believe that our new method 
can serve as a valuable alternative to  
the presently employed techniques, as we do not have to model the functional
form of the HVP in the low-momentum region.

In Sec.~\ref{sect:method} we introduce the analytic continuation method. We then demonstrate in Secs.~\ref{sect:demonstration1} and \ref{sect:demonstration2} the validity of the proposed method using two different approaches.
In Sec.~\ref{sect:Pi}
we perform a computation of the HVP function based on twisted mass fermions, and in Sec.~\ref{sect:amu} we show with the example of $a_\mu^{\rm hvp}$ how the analytic continuation method is used to determine a physical observable.

\section{Alternative method}
\label{sect:method}
In Minkowski spacetime, the HVP function, denoted as $\Pi^{M}(k^2)$, is defined by
\ba
\label{eq:Minkowski}
\Pi^{M}(k^2)(k_\mu k_\nu-g_{\mu\nu}k^2)=i\int d^4x\; e^{-ikx}\;\langle\Omega| T\{ J^{M}_\mu(x) J^{M}_\nu(0)\}|\Omega\rangle\;,
\ea
where $J^{M}_\mu(x)$ is the hadronic component of the electromagnetic vector current
defined in Minkowski spacetime with $x=(x_0,\vec{x})$
and $kx=k_0x_0-\vec{k}\vec{x}$, $k$ is the photon momentum
with $k^2=k_0^2-\vec{k}^2>0$ for timelike and $k^2<0$ for spacelike momenta, and $|\Omega\rangle$ is the QCD vacuum state.

In Euclidean spacetime, the HVP function is given in a similar way
\ba
\label{eq:Euclidean}
\Pi^{E}(K^2)(K_\mu K_\nu-\delta_{\mu\nu}K^2)=\int d^4X\; e^{iKX}\;\langle\Omega| T\{ J^{E}_\mu(X) J^{E}_\nu(0)\}|\Omega\rangle\;,
\ea
with the vector current $J^{E}_\mu(X)$ defined in the Euclidean spacetime with $X=(\vec{x},t)$. 
Here the momenta $K=(\vec{k},K_t)$ are spacelike with $K^2=\vec{k}^2+K_t^2>0$.

By modifying Eq.~(\ref{eq:Euclidean}) we propose to calculate the HVP function in another way:\ 
we perform a Fourier transform only in the spatial directions
but integrate in the time direction with a factor of $e^{\omega t}$, 
\ba
\label{eq:new_method}
\bar{\Pi}(K^2)(K_\mu K_\nu-\delta_{\mu\nu}K^2)=\int dt\;e^{\omega t}\int d^3\vec{x}\;e^{i\vec{k}\vec{x}}\;\langle\Omega| T\{ J^{E}_\mu(\vec{x},t) J^{E}_\nu
(\vec{0},0)\}|\Omega\rangle\;.
\ea
In Eq.~(\ref{eq:new_method}), the momentum $K$ is given by
$K=(\vec{k},-i\omega)$, with $\vec{k}$ the spatial momentum and $\omega$ an input parameter (corresponding here to the photon energy).
By varying $\omega$, we can achieve values for $K^2=-\omega^2+\vec{k}^2$ that enter both
the spacelike and timelike momentum regions. 
In particular, using Eq.~(\ref{eq:new_method}) we can compute $\bar{\Pi}(K^2)$ at zero momentum 
or near $K^2=0.003$ GeV$^2$ without an extrapolation in the momentum.
A very important restriction is that the energy $\omega$ must 
satisfy 
\ba
-K^2=\omega^2-\vec{k}^2<M_V^2\;,\quad\textmd{or}\quad \omega<E_V\;,
\ea
where $E_V$ is the lowest energy in the vector channel
and $M_V$ is the corresponding 
invariant mass.  
Restricting the values of $\omega$ in this way,
a mixing between the photon state and the hadronic states is avoided. Furthermore, 
the divergence caused by $e^{\omega t}$ at infinitely large $t$
is eliminated by a suppression factor $e^{-E_Vt}$ arising from 
the asymptotic time dependence of $\langle J^{E}_\mu(t) J^{E}_\nu(0)\rangle$ thus 
rendering the integral of Eq.~(\ref{eq:new_method}) convergent. 
In the following two sections, we will provide two independent demonstrations that
Eq.~(\ref{eq:new_method}) is the analytic continuation of 
Eq.~(\ref{eq:Minkowski}) from Minkowski spacetime to Euclidean spacetime, and
therefore $\bar{\Pi}(K^2)$ defined in Eq.~(\ref{eq:new_method}) can be used directly
to compute the HVP function and physical quantities where the HVP function enters.

\section{Demonstration using the photon-vector current transition amplitude}
\label{sect:demonstration1}
To demonstrate that Eq.~(\ref{eq:new_method}) is a theoretically valid definition for the HVP function,
we follow Ref.~\cite{Ji:2001wha} closely.
We first consider a matrix element in Minkowski spacetime,
\ba
\label{eq:Minkowski_matrix}
\langle \gamma(k,\lambda)|J^{M}_\nu(0)|\Omega'\rangle\;,
\ea
where $\langle \gamma(k,\lambda)|$ is a photon state with momentum $k$ and 
polarization $\lambda$. It is normalized according to $\langle \gamma(k,\lambda)|\gamma(k',\lambda')\rangle=(2\omega)\delta_{\lambda\lambda'}(2\pi)^3\delta^{(3)}(\vec{k}-\vec{k}')$ 
with $\omega$ the photon energy. 
$|\Omega'\rangle$ denotes the vacuum state in a theory that contains both QCD and QED while $|\Omega\rangle$ is the QCD vacuum state with a trivial QED sector. 

Using the Lehmann$-$Symanzik$-$Zimmermann reduction formula, we convert the matrix element 
in Eq.~(\ref{eq:Minkowski_matrix}) 
into a correlation function,
\ba
\label{eq:LSZ}
\langle \gamma(k,\lambda)|J^{M}_\nu(0)|\Omega' \rangle=
i\lim_{k'\rightarrow k}\varepsilon^{M*}_\mu(k,\lambda)\;{k'}^2\int d^4x\;
e^{-ik'x} \langle\Omega'|T\{A_\mu^{M}(x)J^{M}_\nu(0)\}|\Omega'\rangle\;,
\ea
where $A_\mu^{M}(x)$ is the photon field defined in Minkowski spacetime. It is normalized 
according to $\langle\Omega'|A_\mu^M(x)|\gamma(k,\lambda)\rangle=\varepsilon_\mu^M(k,\lambda)e^{ikx}$, 
and $\varepsilon_\mu^M(k,\lambda)$ is the polarization vector in Minkowski spacetime. 
In perturbative QED, we expand
$\exp(iS_{\rm int})=1+iS_{\rm int}+\cdots$, with $S_{\rm int}=e\int d^4x\;A_{\mu}^{M}(x)J_{\mu}^{M}(x)$, where $e$ is
the electron charge. At $O(e)$
we have 
\ba
\lefteqn{\langle \gamma(k,\lambda)|J^{M}_\nu(0)|\Omega' \rangle =} \nonumber \\ 
&&-e\lim_{k'\rightarrow k}\varepsilon^{M*}_\mu(k,\lambda)\;{k'}^2\int d^4x\;d^4y\;
e^{-ik'x} \langle\Omega'|T\{A_\mu^{M}(x)A_{\rho}^{M}(y)J_{\rho}^{M}(y)J^{M}_\nu(0)\}|\Omega'\rangle\;.
\ea
The Wick contraction of the photon fields $A_\mu^{M}(x)A_{\rho}^{M}(y)$ yields a propagator
$D_{\mu\rho}^M(x,y)=-ig_{\mu\rho}\int \frac{d^4l}{(2\pi)^4}\frac{e^{-il(x-y)}}{l^2+i\varepsilon}$, 
which cancels the inverse propagator ${k'}^2$ outside the integral and results in
\ba
\label{eq:Matrix_Euclidean}
\langle \gamma(k,\lambda)|J^{M}_\nu(0)|\Omega' \rangle=i\;e\;\varepsilon^{M*}_\mu(k,\lambda)\int d^4y\;e^{-iky}\langle\Omega|T\{J_\mu^{M}(y)J^{M}_\nu(0)\}|\Omega\rangle\;.
 \ea

On the other hand, we can define a matrix element $M_{\mu\nu}(t_0,\vec{k})$ in Euclidean spacetime and examine the $t_0\rightarrow\infty$ limit,
\ba
\label{eq:insert_state}
M_{\mu\nu}(t_0,\vec{k})&=&\int d^3\vec{x}\;e^{i\vec{k}\vec{x}}\langle\Omega'|
T\{A^{E}_\mu(\vec{x},t_0)J^{E}_\nu(0)\}|\Omega'\rangle\nn\\
&\rightarrow&\sum_{\lambda}\langle\Omega'|A_{\mu}^{E}(\vec{0},0)|\gamma(k,\lambda)\rangle 
\frac{e^{-\omega t_0}}{2\omega}\langle\gamma(k,\lambda)|J_\nu^{E}(0)|\Omega'\rangle\nn\\
&=&\sum_{\lambda}\varepsilon^E_\mu(k,\lambda)\frac{e^{-\omega t_0}}{2\omega}\langle\gamma(k,\lambda)|J_\nu^{E}(0)|\Omega'\rangle\;,
\ea
where $\omega=|\vec{k}|$ is the on shell photon energy and $A_\mu^E(\vec{x},0)$ is normalized according to $\langle\Omega'|A_\mu^E(\vec{x},0)|\gamma(k,\lambda)\rangle=\varepsilon^E_\mu(k,\lambda)e^{-i\vec{k}\vec{x}}$.
In the second line, we have inserted the ground state $|\gamma(k,\lambda)\rangle$ into 
the matrix element and 
neglected the three-photon states, which are suppressed by powers of the QED coupling. 
We also neglected 
hadronic states because they are suppressed by a factor of $e^{-(E_V-\omega)t_0}$. Here we have used the 
restriction $\omega<E_V$.

The matrix element $M_{\mu\nu}(t_0,\vec{k})$ can be analyzed
by again using leading-order perturbative QED and integrating out the photon field
in Eq.~(\ref{eq:insert_state}). It results in
\ba
M_{\mu\nu}(t_0,\vec{k})=e\int dt\int d^3\vec{y}\; e^{i\vec{k}\vec{y}} D_{\mu\rho}^E(\vec{k},t_0-t)\langle\Omega|T\{J^{E}_\rho(\vec{y},t)J^{E}_\nu(0)\}|\Omega\rangle\;,
\ea
where $D_{\mu\rho}^E(X)=\int \frac{d^4K}{(2\pi)^4}\; e^{iKX} D_{\mu\rho}^E(K)$ is a photon propagator defined in Euclidean spacetime.
Using the integral
\ba
\label{eq:integral}
{\mathcal I}=\int_{-\infty}^{\infty} \frac{dK_t}{2\pi} D_{\mu\nu}^E(K) e^{iK_tt}
=\int_{-\infty}^{\infty} \frac{dK_t}{2\pi} \frac{\delta_{\mu\nu}}{K_t^2+\vec{k}^2} e^{iK_tt}
=\frac{e^{-|\vec{k}||t|}}{2|\vec{k}|}\delta_{\mu\nu}
=\frac{e^{-\omega|t|}}{2\omega}\delta_{\mu\nu}\;,
\ea
we can write
\ba
\label{eq:M_Euclidean}
M_{\mu\nu}(t_0,\vec{k})= e\int dt\;\frac{e^{-\omega|t_0-t|}}{2\omega} \int d^3\vec{y}\; e^{i\vec{k}\vec{y}} \langle\Omega|T\{J_\mu^{E}(\vec{y},t)J_\nu^{E}(0)\}|\Omega\rangle\;.
\ea
We are ultimately interested in the limit of $t_0\rightarrow\infty$
for which $M_{\mu\nu}(t_0,\vec{k})$ becomes
\ba
M_{\mu\nu}(t_0,\vec{k})\rightarrow
e\int dt\;\frac{e^{-\omega (t_0-t)}}{2\omega}\int d^3\vec{y}\; e^{i\vec{k}\vec{y}}
\langle\Omega|T\{J_\mu^{E}(\vec{y},t)J_\nu^{E}(0)\}|\Omega\rangle\,.
\ea
Combining this with Eq.~(\ref{eq:insert_state}) yields
\ba
\label{eq:after_continuation}
\langle\gamma(k,\lambda)|J_\nu^{E}(0)|\Omega'\rangle=
e\;\varepsilon^{E*}_\mu(k,\lambda)\;\int dt\;e^{\omega t}\int d^3\vec{y}\; e^{i\vec{k}\vec{y}}
\langle\Omega|T\{J_\mu^{E}(\vec{y},t)J_\nu^{E}(0)\}|\Omega\rangle\;.
\ea
The left-hand sides of Eqs.~(\ref{eq:Matrix_Euclidean}) and (\ref{eq:after_continuation})
are the matrix elements $\langle \gamma(k,\lambda)|J^{M(E)}_\nu(0)|\Omega' \rangle$, which are equivalent physical observables in Minkowski and Euclidean spacetime.
On the other hand, the right-hand sides of Eqs.~(\ref{eq:Matrix_Euclidean}) and (\ref{eq:after_continuation}) 
are the HVP functions defined
in Eq.~(\ref{eq:Minkowski}) and (\ref{eq:new_method}), respectively. This demonstrates
that these two definitions are equivalent.

In the above analysis, we have looked at the special case that the photon 
is on shell, with $k^2=\omega^2-\vec{k}^2=0$.
In the case that the photon is off shell, we can replace
the matrix element in Eq.~(\ref{eq:Minkowski_matrix}) by
\ba
\langle W(k,\lambda)|J_\nu^M(0)|\Omega'\rangle\;,
\ea
where $W$ denotes a massive vector boson, which has the same quantum 
numbers as the photon but a nonzero mass $M_W$.
Therefore, its momentum $k$ satisfies $k^2=\omega^2-\vec{k}^2=M_W^2>0$.
Using a similar line of arguments as above, we can then show that 
the analytic continuation, as outlined here, is valid for momenta $k^2>0$.
In the next section, we will demonstrate that the method is also valid for $k^2<0$. 

\section{Demonstration from temporal moments}
\label{sect:demonstration2}
In this section we will demonstrate the validity of Eq.~(\ref{eq:new_method})
using a Taylor expansion and the introduction of temporal moments. This
technique has been previously used for lattice~\cite{Allison:2008xk} and perturbative 
calculations~\cite{Kuhn:2007vp,Chetyrkin:2010ic}. 
In Ref.~\cite{Bernecker:2011gh} it has been proposed, to also use it for the calculation
of the HVP function and $a_\mu^{\rm hvp}$.

We write Eq.~(\ref{eq:Euclidean}) as follows:
\ba
\label{eq:Euclidean1}
\Pi^{E}(K^2)F_{\mu\nu}(K)&=&\int dt\;e^{iK_t t} \int d^3\vec{x}\; e^{i\vec{k}\vec{x}}\;\langle J^{E}_\mu(\vec{x},t) J^{E}_\nu(\vec{0},0)\rangle\;,\nn\\
&=&\int dt\;e^{iK_t t}\;C_{\mu\nu}(\vec{k},t)\;,
\ea
where $F_{\mu\nu}(K)=K_\mu K_\nu-\delta_{\mu\nu}K^2$ is the Lorentz factor.

The temporal moments of the correlation function $C_{\mu\nu}(\vec{k},t)$ are computed through
\ba
\label{eq:temp_mom_def}
G_{n,\mu\nu}^{\vec{k}}&\equiv&\frac{(i)^n}{n!} \int dt\; t^n\;C_{\mu\nu}(\vec{k},t)\nn\\
&=&\frac{1}{n!}\frac{\partial^n[\Pi^E(K^2)F_{\mu\nu}(K)]}{\partial (K_t)^n}\Bigg|_{K_t=0}\nn\\
&=&
\left\{
\begin{tabular}{cl}
$M_m^{\vec{k}} F_{\mu\nu}(K)\Bigg|_{K_t=0}
+\frac{1}{2}M_{m-1}^{\vec{k}}\frac{\partial^2 F_{\mu\nu}(K)}{{\partial K_t}^2}\Bigg|_{K_t=0},$ & $\quad$ for $n=2m$,\\
$M_m^{\vec{k}}\frac{\partial F_{\mu\nu}(K)}{\partial K_t}\Bigg|_{K_t=0},$ & $\quad$ for $n=2m+1$,
\end{tabular}\right.
\ea
where $m$ is an integer and the coefficients $M_m^{\vec{k}}$ are given by
\ba
M_m^{\vec{k}}=\frac{1}{m!}\frac{\partial^m \Pi^{E}(K^2)}{\partial(K_t^2)^m}\Bigg|_{K_t=0}\;.
\ea
According to the definition in Eq.~(\ref{eq:temp_mom_def}), the temporal moments $G_{n,\mu\nu}^{\vec{k}}$ are nothing but the coefficients in the Taylor expansion of the function $\Pi^E(K^2)F_{\mu\nu}(K)$ at $K_t=0$.
Using a once-subtracted dispersion relation and the optical theorem
\ba
&&\Pi^E(K^2)-\Pi^E(0)=-\frac{K^2}{\pi}\int ds\;\frac{{\rm Im}[\Pi(s)]}{s(s+K^2)}\;,\nn\\
&&{\rm Im}[\Pi(s)]=\frac{R(s)}{12\pi}\;,\quad R(s)\equiv\frac{\sigma(e^+e^-\rightarrow{\rm hadrons})}{4\pi
\alpha(s)^2/(3s)}\;,
\ea
we can relate $M_m^{\vec{k}}$ to the experimental observables $R(s)$,
\ba
\label{eq:moment_R}
M_{m=0}^{\vec{k}}&=&\Pi^E(0)-\frac{\vec{k}^2}{12\pi^2}\int ds\;\frac{R(s)}{s(s+\vec{k}^2)}\;, \nn\\
M_{m\neq0}^{\vec{k}}&=&(-1)^{m+1}\frac{\vec{k}^2}{12\pi^2}\int ds\; \frac{R(s)}{s(s+\vec{k}^2)^{m+1}}+
(-1)^{m}\frac{1}{12\pi^2}\int ds\; \frac{R(s)}{s(s+\vec{k}^2)^m}\;.
\ea
Note that $M_m^{\vec{k}}$ is suppressed by a factor of $(s+\vec{k}^2)^{-m}\le(M_V^2+\vec{k}^2)^{-m}=(E_V^2)^{-m}$ with $E_V\equiv\sqrt{M_V^2+\vec{k}^2}$ again 
the lowest energy level in the vector channel; see the previous section.
We can construct a convergent series through
\ba
\label{eq:series}
S^{\vec{k}}(\omega^2)=\sum_m M_m^{\vec{k}} (-\omega^2)^m\;,\quad \textmd{if}\;\;\omega^2<M_V^2+\vec{k}^2\;\;
(\omega<E_V)\;.
\ea
Putting Eq.~(\ref{eq:moment_R}) into the series and using
$\Pi^E(0)=\Pi^M(0)$,
we find that $S^{\vec{k}}(\omega^2)$ satisfies the dispersion relation
\ba
S^{\vec{k}}(\omega^2)-\Pi^M(0)=\frac{\omega^2-\vec{k}^2}{12\pi^2}\int ds\;\frac{R(s)}{s\left(s-(\omega^2-\vec{k}^2)\right)}\;,
\ea
which means that $S^{\vec{k}}(\omega^2)$ is equivalent to $\Pi^M(k^2)$ at $k^2=\omega^2-\vec{k}^2$.

On the other hand, we can construct another series through
\ba
S_{\mu\nu}^{\vec{k}}(\omega)&=&\sum_n G_{n,\mu\nu}^{\vec{k}} (-i\omega)^n\nn\\
&=& S^{\vec{k}}(\omega^2)\left(F_{\mu\nu}(K)\Bigg|_{K_t=0}
+(-i\omega)\frac{\partial F_{\mu\nu}(K)}{\partial K_t}\Bigg|_{K_t=0}
+\frac{(-i\omega)^2}{2}\frac{\partial^2 F_{\mu\nu}(K)}{\partial (K_t)^2}\Bigg|_{K_t=0}\right)\nn\\
&=&\Pi^M(k^2)\Bigg|_{k=(\omega,\vec{k})}\;(K_\mu K_\nu-\delta_{\mu\nu}K^2)\Bigg|_{K=(\vec{k},-i\omega)}\;.
\ea
$S_{\mu\nu}^{\vec{k}}(\omega)$ is nothing but $\bar{\Pi}(K^2)(K_\mu K_\nu-\delta_{\mu\nu}K^2)$ as given in
Eq.~(\ref{eq:new_method}). We thus have demonstrated the equivalence 
between $\bar{\Pi}(K^2)$ and $\Pi^M(k^2)$ using temporal moments of the Euclidean 
vector correlation function.

In a special case for $\vec{k}=\vec{0}$ and $\mu=\nu=z$, we have
\ba
G_{0,zz}^{\vec{0}}=0\;,\quad G_{2m+1,zz}^{\vec{0}}=0\;,\quad G_{2m+2,zz}^{\vec{0}}=\frac{1}{2}M_m^{\vec{0}}\neq 0\;.
\ea
The HVP function can then be constructed by
\ba
\label{eq:hvp1}
&&\bar{\Pi}(-\omega^2)=-\frac{1}{\omega^2}\sum_m G_{2m+2,zz}^{\vec{0}}(-i\omega)^{2m+2}=-\int dt\; \frac{e^{\omega t}-1}{\omega^2} C_{zz}(\vec{0},t)\;,\nn\\
&&\bar{\Pi}(-\omega^2)-\bar{\Pi}(0)=-\int dt\; \left[\frac{e^{\omega t}-1}{\omega^2}-\frac{t^2}{2}\right] C_{zz}(\vec{0},t)\;.
\ea
Eq.~(\ref{eq:hvp1}) is the analytic continuation of the formula
\ba
\label{eq:hvp2}
\Pi(K_t^2)-\Pi(0)=\int dt\;\left[\frac{e^{iK_t t}-1}{K_t^2}+\frac{t^2}{2}\right] C_{zz}(\vec{0},t)
\ea
given in Ref.~\cite{Bernecker:2011gh}.

\section{Computation of $\bar{\Pi}(K^2)$}
\label{sect:Pi}
The analytic continuation method described in the previous 
sections has been successfully applied in lattice QCD
calculations of pion and charmonium radiative decay~\cite{Feng:2012ck,Dudek:2006ut}.
In this work we present the first lattice calculation of the HVP function using this
technique.

We will use the same ensembles as in Ref.~\cite{Feng:2011zk}. The gauge configurations
are generated using two-flavor maximally twisted mass fermions~\cite{Baron:2009wt}. The masses of
the up and down quarks are equal and heavier than the physical value with the pion mass
ranging from 650 MeV to 290 MeV. We employ two lattice spacings $a=0.079$ fm and $0.063$ fm 
to check for lattice artifacts and two lattice volumes to examine finite-size effects.

On a finite lattice, Eq.~(\ref{eq:new_method}) takes the form 
\ba
\label{eq:new_method_lattice}
&&\bar{\Pi}(K^2;t_{\rm max})\left(K_\mu K_\nu-\delta_{\mu\nu} K^2\right)=\bar{\Pi}_{\mu\nu}(\vec{k},\omega;t_{\rm max})\;,\nn\\
&&\bar{\Pi}_{\mu\nu}(\vec{k},\omega;t_{\rm max})=\sum_{t=-t_{\rm max}}^{t_{\rm max}-a(\delta_{\mu,t}-\delta_{\nu,t})} e^{\omega (t+a(\delta_{\mu,t}-\delta_{\nu,t})/2)}C_{\mu\nu}(\vec{k},t)\;,
\ea
where $K=(\hat{k}_1,\hat{k}_2,\hat{k}_3,i\hat{\omega})$, with $\hat{k}_i\equiv(2/a)\sin(k_ia/2)$
and $\hat{\omega}\equiv(2/a)\sinh(\omega a/2)$, is the standard lattice definition of four-momentum.
 The correlator $C_{\mu\nu}(\vec{k},t)$ 
is defined by
\ba
\label{eq:correlator}
C_{\mu\nu}(\vec{k},t)=\sum_{\vec{x}} e^{-i\vec{k}(\vec{x}+a\hat{\mu}/2-a\hat{\nu}/2)}\;\langle J^{E}_\mu(\vec{x},t) J^{E}_\nu(\vec{0},0)\rangle\;,
\ea
with $J^{E}_\mu(\vec{x},t)$ the point-split conserved vector current.
The value of $t_{\rm max}$ can be taken up to $T/2$, where $T$ is the temporal extent of the lattice. 

In Eq.~(\ref{eq:new_method_lattice}) we assume that the HVP tensor carries a 
Lorentz factor $K_\mu K_\nu-\delta_{\mu\nu}K^2$.
For some values of $\omega$, we have $K_\mu K_\nu-\delta_{\mu\nu}K^2=0$.
We denote these special values by $\omega_0$.
At $\omega=\omega_0$ and in the large-$t_{\rm max}$ limit, 
the HVP tensor $\bar{\Pi}_{\mu\nu}(\vec{k},\omega;t_{\rm max})$
is supposed to be consistent with zero up to Lorentz symmetry
breaking effects.
In our calculation we do find this to be verified 
for each spatial momentum $\vec{k}$ and polarization direction $\{\mu,\nu\}$.
We therefore calculate $\bar{\Pi}(K^2,t_{\rm max})$ from
\ba
\label{eq:subtract_0}
\bar{\Pi}(K^2;t_{\rm max})=\frac{\bar{\Pi}_{\mu\nu}(\vec{k},\omega;t_{\rm max})-\bar{\Pi}_{\mu\nu}(\vec{k},\omega_0;t_{\rm max})}{K_\mu K_\nu-\delta_{\mu\nu}K^2}\;.
\ea

\subsection{Classification of the correlators}

\begin{figure}[htb]
  \centering
  \includegraphics[trim=15mm 00mm 15mm 00mm, clip, width=280pt,angle=\plotangle]{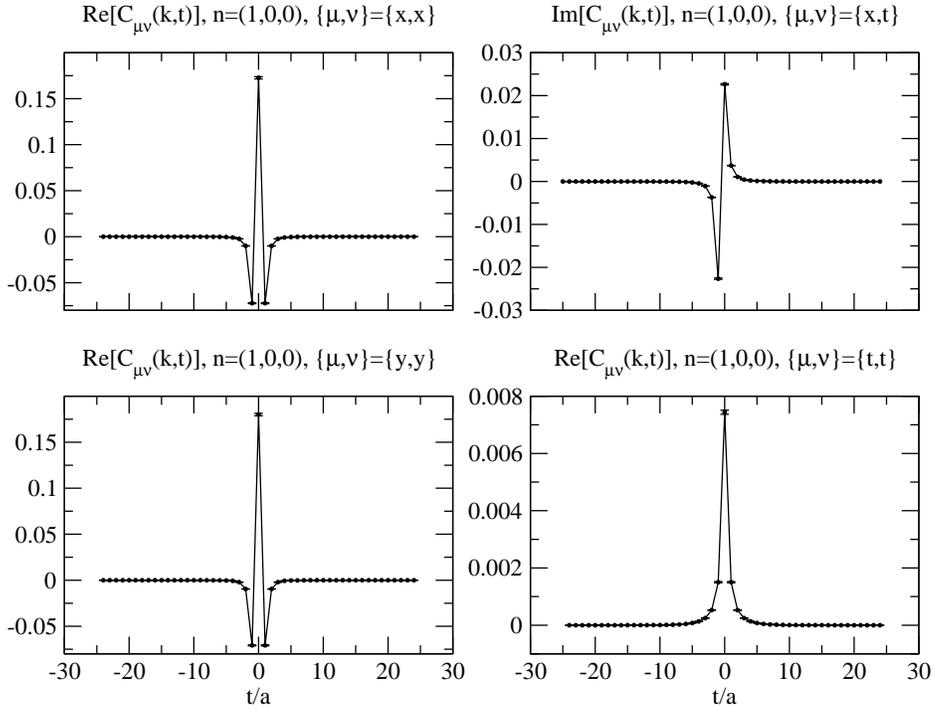}
  \caption{Time dependence of the correlators with different polarization directions.}
  \label{fig:correlator_mode}
\end{figure}

On a lattice with a linear box size $L$, the spatial momenta $\vec{k}$ in Eq.~(\ref{eq:correlator})
take discrete values
\ba
\vec{k}=(2\pi/L)\vec{n}\;,\quad {\rm for}\;\vec{n} \in
\mathbb{Z}^3\;.
\ea
In the calculation performed here, we choose the four lowest momentum modes with 
$\vec{n}=(0,0,0)$, $(1,0,0)$, $(1,1,0)$, and $(1,1,1)$ and will use
$|n|^2$ to distinguish these different modes.

The rotational symmetry indicates that under a (discrete) rotation $\hat{R}\in SO(3,{\mathbb Z})$ with
$SO(3,\mathbb Z)$ the cubic group, the correlators satisfy
\ba
C_{ij}(\hat{R}\vec{k},t)=\Lambda_{ii',jj'}(\hat{R}^{-1}) C_{i'j'}(\vec{k},t)\;,
\quad {\rm for}\;\; i,i',j,j'=x,y,z\;,
\ea
where $\Lambda_{ii',jj'}$ denotes a suitable representation of the group $SO(3,\mathbb Z)$.
In addition to the rotational symmetry, the lattice action is also invariant under parity
and time reversal symmetries~\footnote{Note that the twisted mass fermion action violates parity symmetry. 
In our work we formed a parity average to enforce $O(a)$ improvement.},
which yields~\footnote{Using the local vector current, we will have the relation $C_{\mu\nu}(\vec{k},t)=\eta_{\mu\nu}C_{\mu\nu}(-\vec{k},t)$ and $C_{\mu\nu}(\vec{k},t)=\eta_{\mu\nu}C_{\mu\nu}(\vec{k},-t)$.}
\ba
\label{eq:symmetry}
C_{\mu\nu}(\vec{k},t)=\eta_{\mu\nu}C_{\mu\nu}(-\vec{k},t)\;,\quad
C_{\mu\nu}(\vec{k},t)=\eta_{\mu\nu}C_{\mu\nu}(\vec{k},-t-a(\delta_{\mu,t}-\delta_{\nu,t}))\;,
\ea
with a factor $\eta_{\mu\nu}$ defined by
\ba
\eta_{\mu\nu}=\left\{
\begin{tabular}{l}
$+1,\quad\textmd{for}\;\;\mu,\nu=x,y,z\;,$\\
$-1,\quad\textmd{for}\;\;\mu=t,\;\nu=x,y,z\;\;\textmd{or the reverse}\;,$\\
$+1,\quad\textmd{for}\;\;\mu=\nu=t\;.$
\end{tabular}\right.
\ea

We classify the correlators
using the polarization direction $\{\mu,\nu\}$.
Furthermore, in each class of correlators we average those which are equivalent under the
symmetries of rotation, parity, and time reversal. 
In Table~\ref{tab:subspace}
we list the classification of the correlators
applying the notation of the momentum mode $|n|^2$ and 
the polarization direction $\{\mu,\nu\}$.
\begin{table}[htb]
\centering
\begin{tabular}{|c|c|cccc|ccccc|cccc|}
\hline
$|n|^2$ & $0$ & \multicolumn{4}{c|}{$1$} & \multicolumn{5}{c|}{$2$} & \multicolumn{4}{c|}{$3$} \\
\hline
$\vec{n}$ & $(0,0,0)$ & \multicolumn{4}{c|}{$(1,0,0)$} & \multicolumn{5}{c|}{$(1,1,0)$} & \multicolumn{4}{c|}{$(1,1,1)$} \\
\hline
$\{\mu,\nu\}$ & $\{x,x\}$ 
              & $\{x,x\}$ & $\{x,t\}$ & $\{y,y\}$ & $\{t,t\}$
              & $\{x,x\}$ & $\{x,y\}$ & $\{x,t\}$ & $\{z,z\}$ & $\{t,t\}$
              & $\{x,x\}$ & $\{x,y\}$ & $\{x,t\}$ & $\{t,t\}$ \\
\hline
$N_{\rm sym}$ & $3$
              & $6$ & $12$ & $12$ & $6$
              & $24$ & $24$ & $48$ & $12$ & $12$
              & $24$ & $48$ & $48$ & $8$ \\
\hline
\end{tabular}\caption{Classification of the correlators $C_{\mu\nu}(\vec{k},t)$ with $\vec{k}=(2\pi/L)\vec{n}$. 
$N_{\rm sym}$ is the number of equivalent correlators under rotation, parity and time reversal.}
\label{tab:subspace}
\end{table}

For a given momentum mode $|n|^2$, different polarizations of the vector current can lead
to different types of correlators. Taking $|n|^2=1$ as an example, we show the time dependence
of the correlators $C_{\mu\nu}(\vec{k},t)$ in Fig.~\ref{fig:correlator_mode}.
Notice that, although the correlators themselves are different, 
they lead to the same HVP function, as we will demonstrate below.

\subsection{Finite-size effect}

While Eq.~(\ref{eq:new_method}) requires us to compute an integral in the range from $t=-\infty$ to $t=+\infty$, 
with a given finite lattice volume we can only perform a summation over $t$ values from 
$t=-T/2$ to $+T/2$ as shown in Eq.~(\ref{eq:new_method_lattice}). Thus, 
our calculations are contaminated by finite-size effects, which, however,  
will vanish in the limit $T\rightarrow\infty$.

Ideally, it would be desirable to use the complete $t$ range on the lattice.
However, 
in practice the correlator
$C_{\mu\nu}(\vec{k},t)$ at $|t|$ close to $T/2$ shows very large fluctuations 
and often no useful information can be extracted for these large values
of $|t|$. 
Therefore, we define a maximal $t$ value, $t_{\rm max}=\eta(T/2)$. 
To be concrete, we will set $\eta=3/4$ in the following. Note that, in principle, 
any value of $\eta$ would provide a well-defined choice for our method.  

Of course, on a finite-size lattice the above value of $\eta$ will induce a
finite-size effect. 
This systematic effect is given by 
\ba
\label{eq:thermal}
&&\bar{\Pi}(K^2;t>t_{\rm max})(K_\mu K_\nu-\delta_{\mu\nu}K^2)=
\bar{\Pi}_{\mu\nu}(\vec{k},\omega;t>t_{\rm max})\nn\\
&&\bar{\Pi}_{\mu\nu}(\vec{k},\omega;t>t_{\rm max})\equiv\left(\sum_{t=t_{\rm max}+a-a(\delta_{\mu,t}-\delta_{\nu,t})}^{+\infty}
+\sum_{t=-\infty}^{-t_{\rm max}-a}\right)\; e^{\omega(t+a(\delta_{\mu,t}-\delta_{\nu,t})/2)} C_{\mu\nu}(\vec{k},t) \;.
\ea
In order to obtain an estimate of this finite-size effect, we will assume that for $t>t_{\rm max}$,
the vector correlator is dominated by the ground state. 
We believe that this provides a good estimate of the finite-size effects in our calculation for the following reasons:\ First, for all our ensembles, the contribution
given in Eq.~(\ref{eq:thermal}) is already exponentially suppressed
and thus contributes only little to the total vacuum polarization function. 
Second, even if for $t>t_{\rm max}$ other states may contribute, they provide 
only a correction to a correction and thus should not change our conclusions
significantly. Note that the situation might change if one uses large $\omega$ 
and makes $K^2$ approach the hadron production threshold. 
In this case, the $t>t_{\rm max}$ contribution becomes dominant. 
Besides this, both the energy and the amplitude extracted from $C_{\mu\nu}(\vec{k},t)$ 
are affected by the finite lattice volume. Such effects should be treated properly
using the Lellouch$-$L\"uscher method~\cite{Lellouch:2000pv} and Meyer's proposal 
in Ref~\cite{Meyer:2011um}, which is, however, beyond the scope of this work.

In this calculation we evaluate the HVP function only in the region of spacelike or low timelike momenta. Our strategy will be to 
use as our results the values extracted from $\bar{\Pi}(K^2;t_{\rm max})$ and then
estimate the finite-size effects by computing the contribution of $\bar{\Pi}(K^2;t>t_{\rm max})$. 
It turns out that
with our current lattice setup, the finite-size effects are comparable to 
the statistical error and thus cannot
be neglected in our calculation.
Of course, the calculation 
can be systematically improved 
in future work by using larger volumes and higher statistics.

\subsection{Results for $\bar{\Pi}(K^2)$}

\begin{figure}[htb]
  \centering
\includegraphics[trim=15mm 0mm 15mm 0mm, clip, width=280pt,angle=\plotangle]{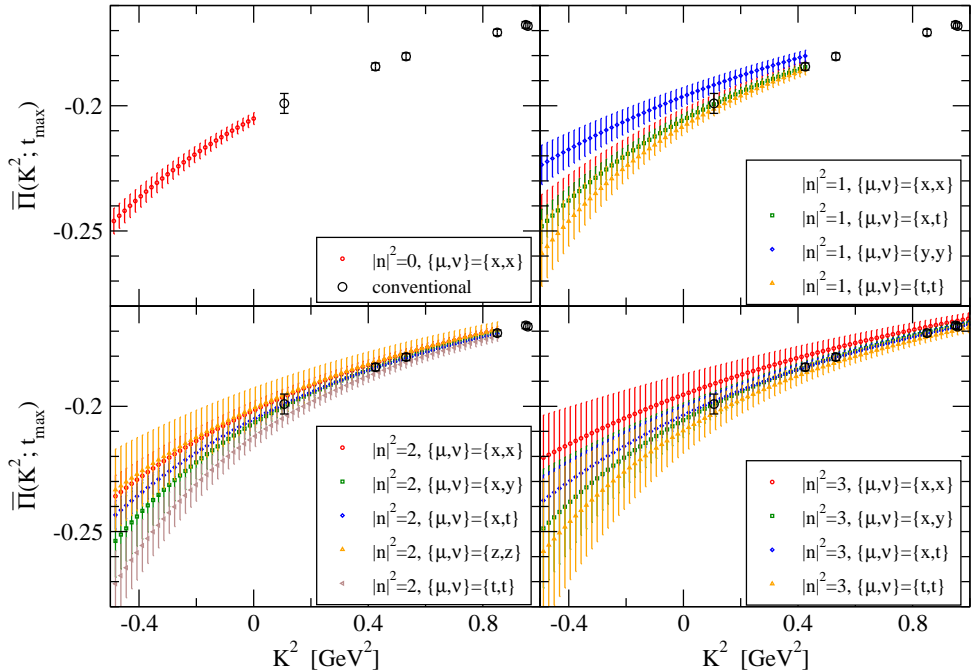}
  \caption{$\bar{\Pi}(K^2;t_{\rm max})$ as a function of $K^2$. 
The ensemble used here features $a=0.079$ fm, $L^3\times T/a^4=24^3\times 48$ and $m_\pi=423$ MeV.}
  \label{fig:Pi_vs_qsq}
\end{figure}

\begin{figure}[htb]
  \centering
\includegraphics[trim=15mm 0mm 15mm 0mm, clip, width=280pt,angle=\plotangle]{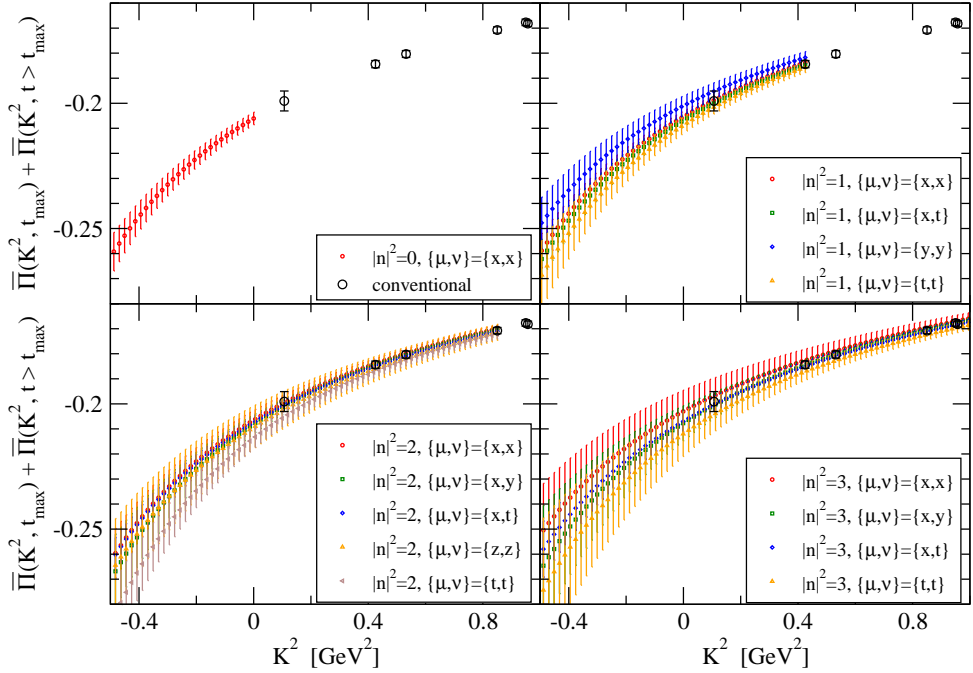}
  \caption{$\bar{\Pi}(K^2;t_{\rm max})+\bar{\Pi}(K^2;t>t_{\rm max})$ as a function of $K^2$.
We use the same ensemble as in Fig.~\ref{fig:Pi_vs_qsq}.
}
  \label{fig:Pi_vs_qsq1}
\end{figure}

\begin{figure}[htb]
  \centering\includegraphics[trim=15mm 0mm 15mm 0mm, clip, width=280pt,angle=\plotangle]{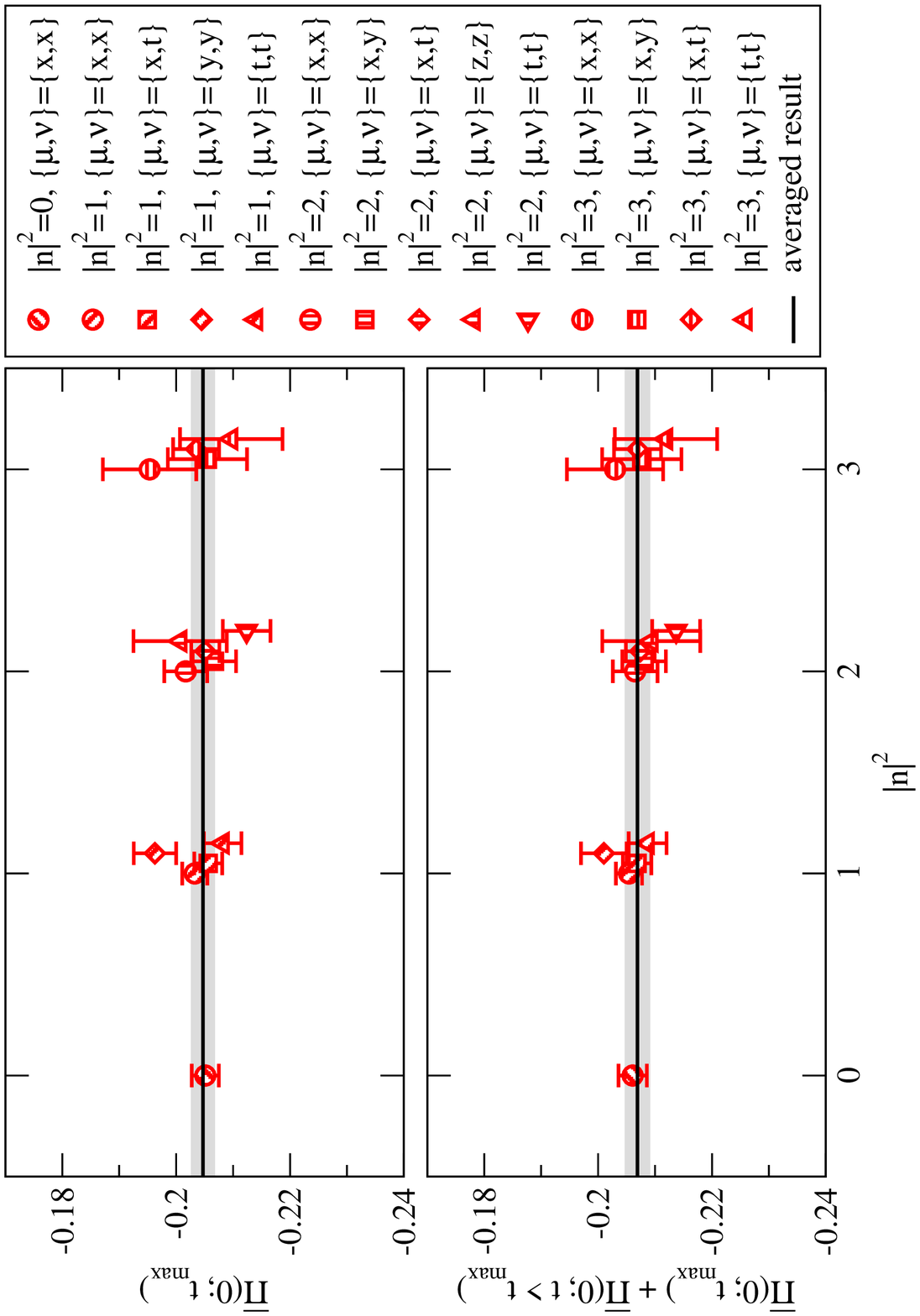}
  \caption{A comparison of $\bar{\Pi}(K^2;t_{\rm max})$ at $K^2=0$
 for different momentum modes $|n|^2$ and polarization directions $\{\mu,\nu\}$. 
In the upper panel, $\bar{\Pi}(0;t_{\rm max})$ is calculated directly using
Eq.~(\ref{eq:new_method_lattice}), while in the lower panel, 
$\bar{\Pi}(0;t_{\rm max})$ is corrected using Eq.~(\ref{eq:thermal}). 
We use the same ensemble as in Fig.~\ref{fig:Pi_vs_qsq}.
Averaging the results for different momentum modes and polarization directions, 
we obtain $\bar{\Pi}(0;t_{\rm max})=-0.2047(20)$ and $\bar{\Pi}(0;t_{\rm max})+\bar{\Pi}(0;t>t_{\rm max})=-0.2069(21)$.}
  \label{fig:Pi0_t_infty}
\end{figure}

We calculate $\bar{\Pi}(K^2;t_{\rm max})$ using Eqs.~(\ref{eq:new_method_lattice}) - (\ref{eq:subtract_0}) 
and show 
$\bar{\Pi}(K^2;t_{\rm max})$ as a function of $K^2$ for different momentum modes
and polarization directions in Fig.~\ref{fig:Pi_vs_qsq}. We demonstrate here that 
using the analytic continuation method, we are able to calculate $\bar{\Pi}(K^2)$ at continuous $K^2$, 
which cover both the spacelike and timelike momentum domain. For some momentum modes, 
e.g., $|n|^2=1$, small discrepancies appear in the HVP functions for different 
polarization directions, which we attribute to finite-size effects. 
As a next step we evaluate these finite-size effects
using Eq.~(\ref{eq:thermal}). The results for 
$\bar{\Pi}(K^2;t_{\rm max})+\bar{\Pi}(K^2;t>t_{\rm max})$ are shown
in Fig.~\ref{fig:Pi_vs_qsq1}. After adding the contribution 
of $\bar{\Pi}(K^2;t>t_{\rm max})$ the results for different polarization directions turn out to be 
more consistent. This finding suggests 
that the discrepancies for $\bar{\Pi}(K^2;t_{\rm max})$ do originate 
from finite-size effects, which then
constitute the dominant systematic effect in our calculation.

Since the HVP functions are consistent among various momentum modes and 
polarization directions, we can average them to obtain the final result. 
Here we perform a weighted average with a weight of $1/\sigma_{\rm stat}^2$, where $\sigma_{\rm stat}$ is the
relative statistical error of $\bar{\Pi}(K^2;t_{\rm max})$ (or $\bar{\Pi}(K^2;t_{\rm max})+\bar{\Pi}(K^2;t>t_{\rm max})$). 
Particularly at $K^2=0$ we show in the upper panel of Fig.~\ref{fig:Pi0_t_infty} the 
averaged result for $\bar{\Pi}(0;t_{\rm max})=-0.2047(20)$ and in the lower 
panel the one for $\bar{\Pi}(0;t_{\rm max})+
\bar{\Pi}(0;t>t_{\rm max})=-0.2069(21)$.
These results deviate at the 1~$\sigma$ level, demonstrating that finite-size effects are 
comparable to the statistical error. 

\begin{figure}[htb]
  \centering\includegraphics[width=340pt,angle=0]{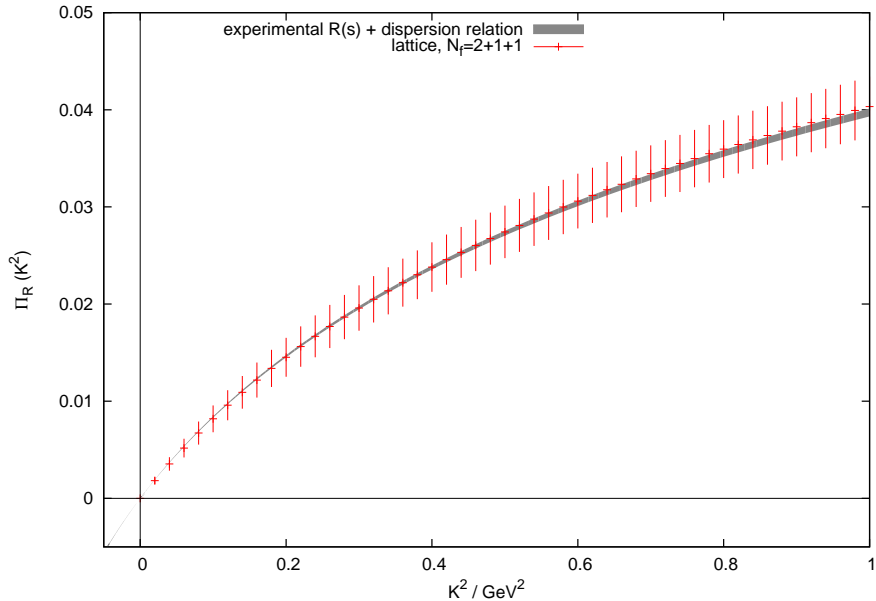}
  \caption{A comparison between $N_f=2+1+1$-flavor lattice results for $\Pi_R(K^2)$ calculated using the analytic continuation method proposed in this work and the experimental results compiled using the cross sections $R(s)$ as input together with the dispersion relation~\cite{Jegerlehner:2011mw}. The lattice results are shown by the red sparse error bars and the experimental data are shown by
the grey condensed error band.
We find that the lattice results are consistent with the experimental data but carry larger errors.}
  \label{fig:comparison_Nf4}
\end{figure}

Up to now, the analysis is performed for the case of $N_f=2$ flavors.
To allow for a direct comparison
with experimental data, we extend the currently used method to 
the case of $N_f=2+1+1$ flavors~\cite{Feng:2012gh}. 
We add the $t>t_{\rm max}$ contribution to the renormalized HVP function $\Pi_R(K^2)=\bar{\Pi}(K^2)-\bar{\Pi}(0)$ 
and extrapolate it to the physical pion mass using the modified extrapolation 
method proposed in Refs.~\cite{Feng:2011zk,Renner:2012fa}. 
In the timelike region, especially the region where $K^2$ approaches the hadron production threshold,  
it is very difficult to reproduce $\Pi_R(K^2)$ due to the significant finite-size effects. 
We therefore restrict the calculation of $\Pi_R(K^2)$ to the spacelike region.
The experimental results for $\Pi_R(K^2)$ are compiled using Jegerlehner's package {\bf alphaQED}~\cite{Jegerlehner:2011mw}, where the dispersion relation is used to relate the experimental data of $R(s)$ (last updated in 2012) to $\Pi_R(K^2)$.
As illustrated in Fig.~\ref{fig:comparison_Nf4}, the lattice results for $\Pi_R(K^2)$ are 
consistent with the experimental data 
but with presently available statistics the corresponding fluctuations are much larger.
Nevertheless, the found
agreement is reassuring that the analytic continuation method can
describe the vacuum polarization function in the low-momentum region.

\section{Determination of $a_\mu^{\rm hvp}$}
\label{sect:amu}
The lattice result of the HVP function obtained in the previous section 
can be used to determine a series of physical observables; see, e.g., Ref.\cite{Renner:2012fa}.
Here we take the leading-order HVP correction to the muon anomalous magnetic moment, 
$a_\mu^{\rm hvp}$, as an
example to study the practical feasibility of the proposed analytic
continuation method.

In lattice QCD $a_\mu^{\rm hvp}$ can be calculated through
\ba
a_\mu^{\rm hvp}=\alpha^2\int_0^{\infty} dK^2\; \frac{1}{K^2}f\left(\frac{K^2}{m_\mu^2}\right)(\Pi(K^2)-\Pi(0))\;,
\ea
where $\alpha$ is the fine structure constant, 
$m_\mu$ is the muon mass, and $f(K^2/m_\mu^2)$ is a known function~\cite{Blum:2002ii}, 
which assumes a maximum at $K^2=(\sqrt{5}-2)m_\mu^2\approx 0.003$ GeV$^2$.
To control the chiral extrapolation, we use a modified definition proposed in Refs.~\cite{Feng:2011zk,Renner:2012fa},
\ba
\label{eq:amu}
a_{\bar{\mu}}^{\rm hvp}=\alpha^2\int_0^{\infty} dK^2\; \frac{1}{K^2}
f\left(\frac{K^2}{m_\mu^2}\frac{H_{\rm phys}^2}{H^2}\right)(\Pi(K^2)-\Pi(0))\;,
\ea
with $H=M_V$. Note that when the pion mass approaches its physical value, we have 
$H=H_{\rm phys}$. Thus, the modified definition, $a_{\bar{\mu}}^{\rm hvp}$,
reproduces the value of $a_\mu^{\rm hvp}$ at the physical pion mass.   

In conventional lattice calculations, to perform the integral in Eq.~(\ref{eq:amu}), one needs to
parametrize the HVP function. 
Using the analytic continuation method, we can calculate the HVP function for a 
continuous momentum region
$0<K^2<K_{\rm max}^2$, with $K_{\rm max}^2=\sum_{i=x,y,z}\hat{K}_i^2$ 
being the squared spatial momentum.
Thus, we can avoid the parametrization in this momentum region. In practice
we split
 Eq.~(\ref{eq:amu}) into three parts:
\ba
\label{eq:amu_split}
a_{\bar{\mu}}^{\rm hvp}&=&a_{\bar{\mu}}^{(1)}+a_{\bar{\mu}}^{(2)}+a_{\bar{\mu}}^{(3)}\;,\nn\\
a_{\bar{\mu}}^{(1)}&=&\alpha^2\int_0^{K_{\rm max}^2} dK^2\; \frac{1}{K^2}
f\left(\frac{K^2}{m_\mu^2}\frac{H_{\rm phys}^2}{H^2}\right)(\Pi(K^2)-\Pi(0))\;,\nn\\
a_{\bar{\mu}}^{(2)}&=&\alpha^2\int_{K_{\rm max}^2}^{\infty} dK^2\; \frac{1}{K^2}
f\left(\frac{K^2}{m_\mu^2}\frac{H_{\rm phys}^2}{H^2}\right)(\Pi(K_{\rm max}^2)-\Pi(0))\;,\nn\\
a_{\bar{\mu}}^{(3)}&=&\alpha^2\int_{K_{\rm max}^2}^{\infty} dK^2\; \frac{1}{K^2}
f\left(\frac{K^2}{m_\mu^2}\frac{H_{\rm phys}^2}{H^2}\right)(\Pi(K^2)-\Pi(K_{\rm max}^2))\;.
\ea
In Eq.~(\ref{eq:amu_split}), $a_{\bar{\mu}}^{(1)}+a_{\bar{\mu}}^{(2)}$
can be calculated directly using the analytic continuation method.
As shown by Eq.~(\ref{eq:new_method_lattice}),   
$\bar{\Pi}(K^2;t_{\rm max})$ is a linear combination of the correlator $C_{\mu\nu}(\vec{k},t)$. 
Putting this definition of $\bar{\Pi}(K^2;t_{\rm max})$ into Eq.~(\ref{eq:amu_split}), 
it turns out that
 $a_{\bar{\mu}}^{(1)}$ and $a_{\bar{\mu}}^{(2)}$ are also linear combinations of 
$C_{\mu\nu}(\vec{k},t)$ with coefficients that can be determined by performing 
the integral in Eq.~(\ref{eq:amu_split}). 
Similar to the previous section, 
we can calculate $a_{\bar{\mu}}^{(1)}$ and $a_{\bar{\mu}}^{(2)}$ up to 
the value of $t_{\rm max}=\eta(T/2)$ with $\eta=3/4$ and estimate the 
finite-size effects using $\bar{\Pi}(K^2;t>t_{\rm max})$.

The evaluation of $a_{\bar{\mu}}^{(3)}$ still requires a parametrization of $\bar{\Pi}(K^2)$. 
This will bring in some model dependence in our analysis, which, however,
is a small effect, since the total contribution of $a_{\bar{\mu}}^{(3)}$ only 
amounts to a few percent in case of momentum modes $|n|^2=1,2,3$.
The parametrization of $\bar{\Pi}(K^2)$ used in this calculation is given in the Appendix.

In Fig.~\ref{fig:corr_amu} we show $a_{\bar{\mu}}^{\rm hvp}$ as a function of the squared pion mass.
The results of $a_{\bar{\mu}}^{\rm hvp}$ calculated using Eq.~(\ref{eq:amu_split}) are shown
by the empty symbols. These results have been averaged among various momentum modes ($|n|^2=1,2,3$) and polarization directions.
For comparison the results determined using our conventional approach of
parametrizing the HVP function in the full momentum range are shown in
the same figure, represented by the filled symbols.
In the upper panel we show the results of 
$a_{\bar{\mu}}^{\rm hvp}(t_{\rm max})$, which are calculated using the correlator $C_{\mu\nu}(\vec{k},t)$ covering the range of
$-t_{\rm max} \le t\le t_{\rm max}$. We look at the finite-size effects 
in $a_{\bar{\mu}}^{\rm hvp}(t_{\rm max})$
by comparing the results for different volumes. There are also some deviations between the 
results from the analytic continuation method and the standard parametrization method.
To check for the finite-size effects, we evaluate the contribution to $a_{\bar{\mu}}^{\rm hvp}$ 
from $C_{\mu\nu}(\vec{k},t)$ at $|t|>t_{\rm max}$ leading to a correction
$a_{\bar{\mu}}^{\rm hvp}(t>t_{\rm max})$. The corresponding results for 
$a_{\bar{\mu}}^{\rm hvp}(t_{\rm max})+a_{\bar{\mu}}^{\rm hvp}(t>t_{\rm max})$ 
are shown in the lower panel of Fig.~\ref{fig:corr_amu}. These results are consistent now
among different lattice
volumes. Besides this, the results from the analytic continuation method also agree with the ones from
the standard parametrization method. 
As can be seen, 
the results from the 
analytic continuation method for $a_{\bar{\mu}}^{\rm hvp}$ 
show larger fluctuations than the standard ones. However, the analytic 
continuation method has the conceptual advantage that  
in the region of low $K^2$ the parametrization of
$\bar{\Pi}(K^2)$ can be avoided.
Thus, we think that presently the analytic
continuation method can serve as a valuable cross-check of the standard
method to analyze the vacuum polarization function. 
\begin{figure}[ht]
\begin{center}
  \includegraphics[trim=10mm 00mm 15mm 00mm, clip, width=280pt,angle=\plotangle]{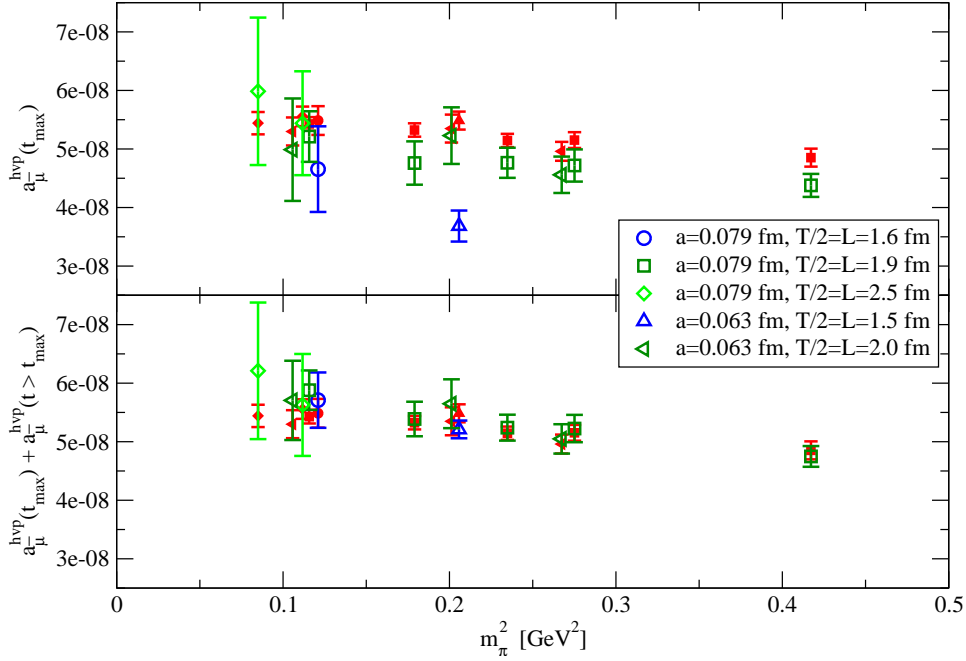}
\end{center}
\caption{$a_{\bar{\mu}}^{\rm hvp}$ as a function of the squared pion mass. For each ensemble, the empty symbols
stand for the results calculated using the analytic continuation method.
We have averaged $a_{\bar{\mu}}^{\rm hvp}$ for various momentum modes ($|n|^2=1,2,3$) and polarization directions.
In the upper panel, we show $a_{\bar{\mu}}^{\rm hvp}(t_{\rm max})$, which are calculated using
the correlator $C_{\mu\nu}(\vec{k},t)$ covering the range $-t_{\rm max}\le t\le t_{\rm max}$.
In the lower panel, we show the total contribution of 
$a_{\bar{\mu}}^{\rm hvp}(t_{\rm max})+a_{\bar{\mu}}^{\rm hvp}(t>t_{\rm max})$, where an estimate of the finite-size effects has been added.
The filled symbols indicate the results for $a_{\bar{\mu}}^{\rm hvp}$ 
from the conventional parametrization method.} 
\label{fig:corr_amu}
\end{figure}

\section{Conclusions}

The analytic continuation method, proposed in this 
work, allows us to obtain information on the momentum dependence 
of the vacuum polarization function at small momenta, which is a significant
challenge for lattice QCD calculations. 
In order to see how the analytic continuation 
method works in practice, we performed a pilot
study for 
the HVP function  
to determine the 
leading-order QCD correction to the muon anomalous magnetic moment. 
Since at large Euclidean times the HVP is very noisy,  
we restricted the time summation to a maximum 
time, $t_{\rm max}=\eta T/2$, with 
$T$ the time extent of our lattice. In this work, we have chosen 
$\eta=3/4$. Although in the infinite volume limit this would 
lead to a fully correct definition for $a_\mu^{\rm hvp}$, on a finite
lattice such a choice induces a finite-size effect. We estimated 
this finite-size effect by assuming that in the time region 
excluded by the cut $\eta=3/4$ the ground state dominates 
in the vector channel. 

In the case of $N_f=2+1+1$ flavors, adding the so-computed 
finite-size effects to the results for the HVP function
provides a result that agrees with the experimental 
determination of Ref.~\cite{Jegerlehner:2011mw}. 

Going to the case of $N_f=2$ flavors, we find after the finite-size correction, 
consistent 
results for $a_\mu^{\rm hvp}$ for all ensembles and also an agreement 
with results computed earlier by us using the conventional 
approach. We therefore conclude that applying a cut in 
the Euclidean time induces indeed a finite-size effect as 
the dominant systematic error. Thus, when the analytic 
continuation method is applied on larger lattices in the future,
any assumption such as the here employed ground state dominance 
at large times can be completely avoided, leading to 
a conceptually clear determination of quantities 
derived from the HVP function.  
In this
way it would not be required anymore to rely on model dependent
parametrizations that enter some of the conventional
computations of $a_\mu^{\rm hvp}$. 
On the negative side, 
it needs to be said that, although having a conceptual 
advantage, the analytic continuation method 
gives results, at least for $a_\mu^{\rm hvp}$, which have  
fluctuations that are larger than the conventional method. 

However, 
since all the methods are afflicted with different 
systematic uncertainties, comparing results from different approaches will 
provide confidence in the extraction of physical quantities where the 
HVP function is an essential ingredient. 
In this sense we believe that 
the analytic continuation procedure
presented in this paper
can provide a valuable alternative to other approaches to determine the HVP function.
It will be interesting to test the potential 
of the here proposed analytic continuation method 
for quantities different from the HVP function, where the momentum 
dependence at small or even zero momentum is not directly accessible. 

\begin{acknowledgments}
X.F. and S.H. are supported
in part by the Grant-in-Aid of the Japanese Ministry of Education (Grant No.\ 21674002), and
D.R. is supported in part by Jefferson Science Associates, LLC, under 
U.S. DOE Contract No.\ DE-AC05-06OR23177.
G.H. gratefully acknowledges the support of the German Academic National Foundation 
(Studienstiftung des deutschen Volkes e.V.)
and of 
the DFG-funded corroborative
research center SFB/TR9.
K.J. is supported in part by the Cyprus Research Promotion
Foundation under Contract No. $\Pi$PO$\Sigma$E$\Lambda$KY$\Sigma$H/EM$\Pi$EIPO$\Sigma$/0311/16.
The numerical computations have been performed on the
{\it SGI system HLRN-II} at the {HLRN Supercomputing Service Berlin-Hannover},  FZJ/GCS,
and BG/P at FZ-J\"ulich.
The analysis was performed on the computer centers of KEK, DESY Zeuthen, and Humboldt
University Berlin.
\end{acknowledgments}

\appendix
\section{Parametrization of the HVP function}
In the conventional approach, in order to
compute the integral in Eq.~(\ref{eq:amu}), we need a functional form
to describe the $K^2$ dependence of $\Pi(K^2)$.
 For low $K^2$, we use the form
\ba
\Pi_{\rm low}(K^2) = \frac{- g_V^2 M_V^2}{K^2 + M_V^2} + \left(a_0 + a_1 K^2\right)\;,
\ea
where the first term is the dominant contribution from the ground-state vector
meson and $g_V$ is the electromagnetic coupling of the vector meson.
The polynomial terms account for residual contributions.
$\Pi(0)$ is given by
$-g_V^2+a_0$.
For high $K^2$, we use the form
\ba
\Pi_{\rm high}(K^2) = c + \ln(K^2)\left(
\sum_{i=0}^{3} b_i (K^2)^i \right)\;,
\ea
with a free constant $c$ and a sum over four terms in the polynomial multiplying the logarithmic function.
We combine the two expressions using
\ba
\label{eq:Pi_total}
\Pi(K^2) = \frac{1-t}{2}\Pi_{\rm low}(K^2) + \frac{1 + t}{2}\Pi_{\rm high}(K^2)\;,
\ea
where $t = {\rm {tanh}}( ( K^2 - K_m^2 ) / \Delta_m^2 )$ is a smooth approximation to the step function.
The parameters
are set as $K_m = 1.3 M_V$ and $\Delta_m = 0.3 M_V$.
We fit the lattice data of $\Pi(K^2)$ to Eq.~(\ref{eq:Pi_total}) with seven free parameters:
$a_0$, $a_1$, $c$, $b_0$, $b_1$, $b_2$ and $b_3$.

\bibliography{hvp}
\bibliographystyle{h-physrev}

\end{document}